\documentclass[reprint, amsmath, amssymb, aps, superscriptaddress]{revtex4-1}
\usepackage{format}
\usepackage{soul} 
\allowdisplaybreaks

\begin{document}

\preprint{APS/123-QED}


\title{ Bound-state-free F\"orster resonant shielding of strongly dipolar ultracold molecules }

\author{Reuben R. W. Wang}
\affiliation{ ITAMP, Center for Astrophysics $|$ Harvard \& Smithsonian, Cambridge, Massachusetts 02138, USA }
\affiliation{ Department of Physics, Harvard University, Cambridge, Massachusetts 02138, USA }

\date{\today} 

\begin{abstract}

We propose a method to suppress collisional loss in strongly dipolar, rotationally excited ultracold molecules using a combination of static (dc) and microwave (ac) electric fields. 
By tuning two excited pair molecular rotational states into a F\"orster resonance with a dc field, simultaneously driving excited rotational transitions with an ac field removes all long-range bound states, allowing near complete suppression of all two- and three-body collisional loss channels.  
While permitting tunable dipolar and anti-dipolar interactions, this bound-state-free ac/dc scheme is not subject to photon-changing collisions that are the primary source of two-body loss in shielding with two microwave fields, used to achieve the first molecular Bose-Einstein condensate [Bigagli et al., Nature 631, 289 (2024)].
Using NaCs as a representative example for strongly dipolar molecules, close-coupling calculations are performed to show that bound-state-free shielding can achieve ratios of elastic-to-loss rates $\gtrsim 10^{6}$ at 100 nK, with currently accessible ac and dc field generation technologies. This work opens new opportunities for realizing large, long-lived samples of strongly interacting degenerate molecular gases with tunable long-range interactions.    

\end{abstract}

\maketitle

\textit{Introduction}--
The collisional shielding of ultracold molecules with static (dc) electric \cite{Avdeenkov06_PRA, Wang15_NJP, Quemener16_PRA, Matsuda20_Sci} and microwave (ac) fields \cite{Cooper09_PRL, Karman18_PRL, Lassabliere18_PRL, Anderegg21_Sci} have been crucial in achieving stable quantum gases of polar molecules \cite{Li21_Nat, Schindewolf22_Nat, Bigagli23_NatPhys, Lin23_PRX, Schindewolf25_arxiv}.  
In strongly dipolar diatomic molecules such as NaCs and KAg, the long-range intermolecular potential from dc field shielding hosts weakly-bound tetratomic states \cite{Mukherjee24_PRR}, discovered and coined by Avdeenkov and Bohn as field-linked (FL) states \cite{Avdeenkov03_PRL}. 
The presence of FL states can be harnessed to tune the two-body scattering length in dipolar molecules \cite{Lassabliere18_PRL, Mukherjee24_PRR}, but also results in additional three-body recombination loss channels in dense gases \cite{Stevenson24_PRL}, limiting the lifetimes and achievable densities of bulk molecular samples through evaporative cooling \cite{Bigagli23_NatPhys, Lin23_PRX}.

Recently, collisional shielding with a combination of circularly and linearly polarized microwaves was employed to eliminate all FL states \cite{Karman25_PRXQ}, allowing large suppression of both two and three-body losses \cite{Yuan25_arxiv}. 
This double microwave shielding technique has now allowed the realization of Bose-Einstein condensation of groundstate NaCs \cite{Bigagli24_Nat} and NaRb \cite{Shi25_arxiv} molecules. 
Unfortunately, the double microwave scheme is subject to Floquet inelastic processes, where the two microwave fields exchange photons during a molecular collision, transferring a beat frequency worth of kinetic energy to the molecules \cite{Stevenson26_inprep}. 
These photon-changing collisions are, in fact, the dominant loss-channel in double microwave shielded molecules \cite{Karman25_PRXQ}, motivating alternative shielding schemes that achieve similar interaction potentials.

In this work, we propose a shielding protocol that eliminates both FL states and photon-changing losses with current-day accessible technical requirements. 
Our scheme shares similarities with an earlier proposal by Gorshkov et al. \cite{Gorshkov08_PRL} with groundstate molecules, but additionally leverages the higher two-body loss suppression achievable through dc field shielding with excited rotational states \cite{Quemener16_PRA, GonzalezMartinez17_PRA, Mukherjee24_PRR}. An ac field then provides additional near-resonant couplings to other rotational states, allowing complete elimination of all FL states while improving the efficacy of collisional shielding, the details of which we now describe.

\begin{figure}[ht]
    \centering
    \includegraphics[width=\linewidth]{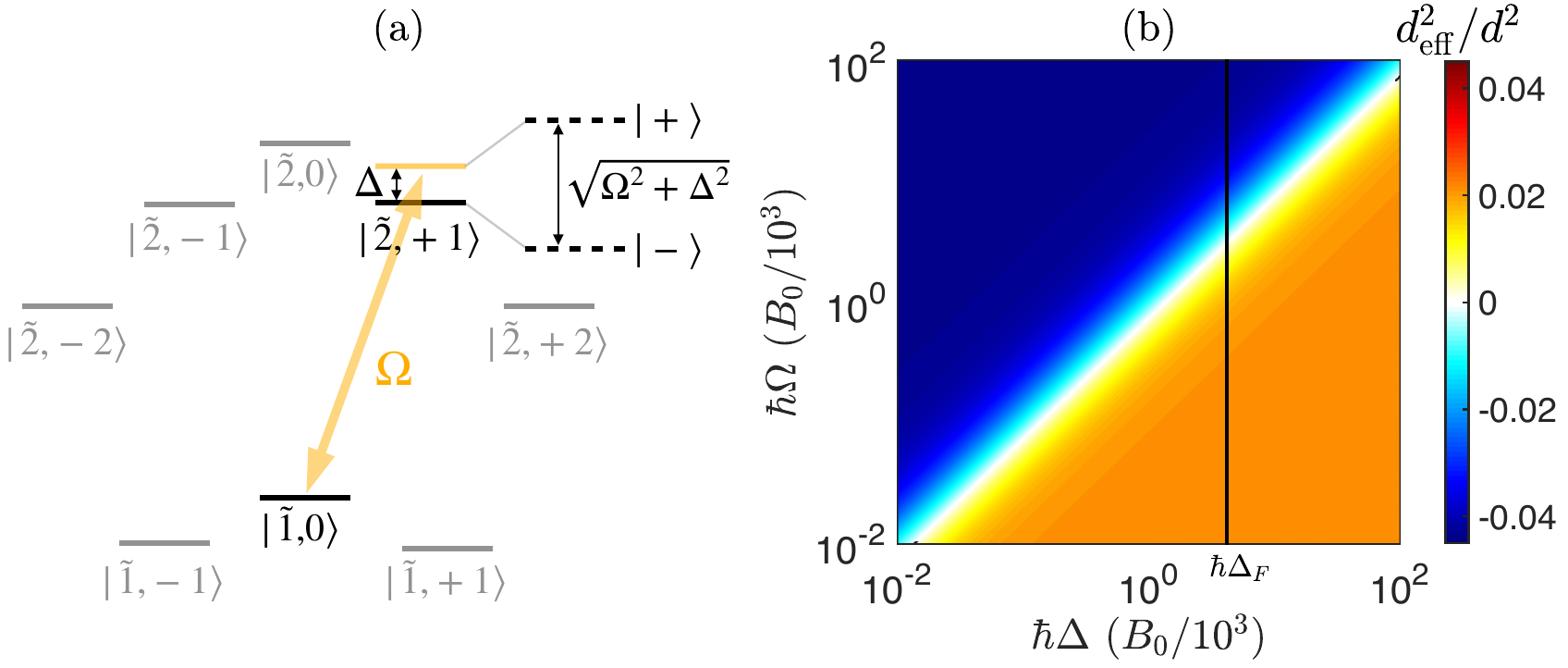}
    \vspace{-20pt}
    \caption{ (a) Illustration of the ac/dc dressing scheme for bound-state-free shielding. (b) The squared effective dipole moment as a function of ac detuning and Rabi frequency. White on the colorbar indicates a zero effective dipole moment, while the solid black vertical line indicates the F\"orster defect along the ac detuning axis.  }
    \label{fig:ACDC_shielding_scheme_summary}
\end{figure}

\textit{Shielding scheme}--
Our scheme requires applying a dc field large enough to overcome the rotational splittings determined by the rotational constant $B_0$, shifting the molecular rotational levels labeled by the rotational quantum numbers $\ket{ N, m }$. At a specific field strength, the pair molecular state $| \tilde{N}_A, m_{A}; \tilde{N}_B, m_{B}\rangle_S = | \tilde{1}, 0; \tilde{1}, 0\rangle_S$, is brought close to a F\"orster resonance with $| \tilde{0}, 0; \tilde{2}, 0\rangle_S$, with the former pair state placed slightly above the latter by energy $\hbar\Delta_F$, referred to as the F\"orster defect. 
Tildes are used to denote dc field-dressed quantum numbers, and a subscript $S$ indicates that the state has been symmetrized under exchange of molecules $A$ and $B$.
Two such dc dressed molecules prepared in $| \tilde{1}, 0; \tilde{1}, 0\rangle_S$ then experience resonant dipole-dipole interactions (DDI) upon close approach, where the field dressing turns nominally attractive rotational van der Waals interactions repulsive.  
F\"orster resonant shielding is expected to be more effective in diatomic molecules with larger rotational constants $B_0$ and molecular-frame dipole moments $d$ \cite{GonzalezMartinez17_PRA, Mukherjee24_PRR}.

In addition to the dc field, we also drive the $|{\tilde{N}, m}\rangle = |\tilde{1}, 0\rangle \rightarrow |\tilde{2}, +1\rangle$ transition with a circularly polarized microwave of Rabi frequency $\Omega$ and blue-detuning $\Delta$. 
We consider the regime where the ac field strength ${\rm E}_{\rm ac}$ satisfies $d{\rm E}_{\rm ac} \ll B_{0}$, as is accessible in table-top molecular experiments. 
This condition limits the achievable magnitude of $\Omega$, and correspondingly the allowable $\Delta$ to permit an appreciable ac Stark effect, which are typically much smaller than $\Delta_F$ at the optimal dc shielding field of $d{\rm E}_{\rm dc} \approx 3.4 B_0$ \cite{GonzalezMartinez17_PRA}.
Instead, our scheme utilizes a slightly lower dc field of $d{\rm E}_{\rm dc} = 3.25 B_0$, just above the resonant crossing point of $d{\rm E}_{\rm dc} \approx 3.244 B_0$, so that $\Omega, \Delta$ and $\Delta_F$ are all comparable in magnitude.
The corresponding F\"orster energy defect and Rabi frequency are then $\hbar\Delta_F \approx 4.17 \times 10^{-3} B_0$ and $\Omega \approx 0.4577 d{\rm E}_{\rm ac}/\hbar$ respectively.  
Although this field is sub-optimal for pure F\"orster resonant dc shielding, strongly dipolar molecules such as NaCs or KAg can still achieve $\gtrsim 10^{6}$ order of magnitude elastic-to-loss collision rate ratio \cite{Mukherjee24_PRR}, which can be further increased with the added ac field to allow highly efficient collisional shielding.      
We focus on NaCs molecules in this work as a representative candidate for bound-state-free ac/dc shielding.
At the chosen dc field value, NaCs inherits a F\"orster defect of $\Delta_F/(2\pi) \approx 7.25$ MHz, while the shielded adiabatic potential supports a shallow FL state in the absence of an ac field.

The effect of applying both dc and ac fields on a rigid-rotor molecule is visualized in subplot (a) of Fig.~\ref{fig:ACDC_shielding_scheme_summary}, illustrating that the dc field sets a background rotational structure on which the ac field is used to further tune intermolecular interactions. 
In the rotating frame of the microwaves, the ac drive mixes the two bright states (black solid lines in Fig.~\ref{fig:ACDC_shielding_scheme_summary}\textcolor{blue}{a}) into upper and lower $\ket{ \pm }$ dressed states (black dashed lines in Fig.~\ref{fig:ACDC_shielding_scheme_summary}\textcolor{blue}{a}) with respective eigenenergies $\varepsilon_{\pm} = -\hbar\Delta/2 \pm \hbar\sqrt{ \Delta^2 + \Omega^2 } / 2$.  
As the two-molecules approach, they interact at long-range primarily via DDI. 
The nature of two-molecule interactions under the dressing scheme described above is most easily understood through the effective adiabatic potential energy surface that they collide elastically upon. Such an effective potential is obtained with a rotating wave approximation, restricting the Hilbert space of two rigid rotors to the exchange symmetric pair basis set:
\begin{align} \label{eq:DCdressed_symmetrized_basis}
    \begin{tabular}{l|l|c}
        state label & $|{ \tilde{N}_A, m_{N_A}; \tilde{N}_B, m_{N_B} }\rangle_S\ket{ n }$ & energy \\
        \hline
        $|g; g\rangle\ket{0}$ & $|{ \tilde{1},0; \tilde{1},0 }\rangle_S\ket{ 0 }$ & $0$ \\
        $|F\rangle\ket{0}$ & $|{ \tilde{2}, 0; \tilde{0}, 0 }\rangle_S\ket{ 0 }$ & $-\hbar\Delta_{F}$  \\
        $|\bar{e}; g\rangle\ket{-1}$ & $|{ \tilde{2}, -1; \tilde{1}, 0 }\rangle_S\ket{ -1 }$ & $-\hbar\Delta$ \\
        $|e; g\rangle\ket{-1}$ & $|{ \tilde{2}, +1; \tilde{1}, 0 }\rangle_S\ket{ -1 }$ & $-\hbar\Delta$ \\
        $|e; \bar{e}\rangle\ket{-2}$ & $|{ \tilde{2}, +1; \tilde{2}, -1 }\rangle_S\ket{ -2 }$ & $-2 \hbar\Delta$ \\
        $|e; e\rangle\ket{-2}$ & $|{ \tilde{2}, +1; \tilde{2}, +1 }\rangle_S\ket{ -2 }$ & $-2 \hbar\Delta$
    \end{tabular},
\end{align}
where $\ket{ n }$ is the Fock state for differential microwave photon number $n$. 
Within this restricted subspace, the shielded adiabatic surface is derived to be: 
\begin{align} \label{eq:effective_potential}
    V_{\rm eff}(\boldsymbol{r})
    &=
    V^{(1)}(\boldsymbol{r})
    +
    V^{(2)}_{\rm dc}(\boldsymbol{r})
    +
    V^{(2)}_{\rm ac}(\boldsymbol{r}) \nonumber\\
    &=
    \frac{ d_{\rm eff}^2 }{ 4 \pi \epsilon_0 }
    \frac{ 1 - 3 \cos^2\theta }{ r^3 }
    +
    V^{(2)}_{\rm dc}(\boldsymbol{r})
    +
    V^{(2)}_{\rm ac}(\boldsymbol{r}),
\end{align}
with effective squared dipole moment $d_{\rm eff}^2 = ( \alpha_{-} d_{e} - \alpha_{+} d_g )^2 - \alpha_{-} \alpha_{+} d_{g\rightarrow{e}} d_{{e}\rightarrow{g}}$ given in terms of the coefficients $\alpha_{\pm} = \{ [ 1 + 2 \delta \left(\delta \pm \sqrt{\delta^2 + 1}\right) ] / ( 4\delta^2 + 4 ) \}^{1/2}$, ratio $\delta = \Delta/\Omega$, second-order potentials
\begin{subequations} \label{eq:secondorder_potential}
\begin{align}
    V^{(2)}_{\rm dc}(\boldsymbol{r})
    &=
    \left(
    \frac{ d_{\tilde{1}\rightarrow\tilde{2}}^{0\rightarrow 0} d_{\tilde{1}\rightarrow\tilde{0}}^{0\rightarrow 0} }{ 4 \pi \epsilon_0 r^3 }
    \right)^2
    \frac{ 2 \alpha_{+}^2 (1 - 3\cos^2\theta)^2 }{ 2\varepsilon_{+} + \hbar\Delta_F }, \\
    V^{(2)}_{\rm ac}(\boldsymbol{r})
    &=
    \left(
    \frac{ d_{g \rightarrow e} d_{\bar{e} \rightarrow g} }{ 4 \pi \epsilon_0 r^3 }
    \right)^2
    \frac{ \alpha_+^2 - \alpha_+ \alpha_- }{ 8 \delta \sqrt{ 1 + \delta^2 } }
    \frac{ 9\sin^4\theta }{ \varepsilon_+ + \hbar\Delta },
\end{align}
\end{subequations}
direct dipole moments $d_{g}, d_{e}, d_{\bar{e}}$, and transition dipole moments $d_{g\rightarrow{e}}, d_{e\rightarrow{g}}, d_{\bar{e}\rightarrow{g}}, d_{\tilde{1}\rightarrow\tilde{2}}^{0\rightarrow 0}, d_{\tilde{1}\rightarrow\tilde{0}}^{0\rightarrow 0}$.
Above, $\theta$ is the angle between $\boldsymbol{r}$ and the effective lab-frame dipole moment. 
Details of the derivation and explicit definitions of the dipole moments are provided in App.~\ref{app:effective_potential_derivation}. 

\begin{figure}[ht]
    \centering
    \includegraphics[width=\linewidth]{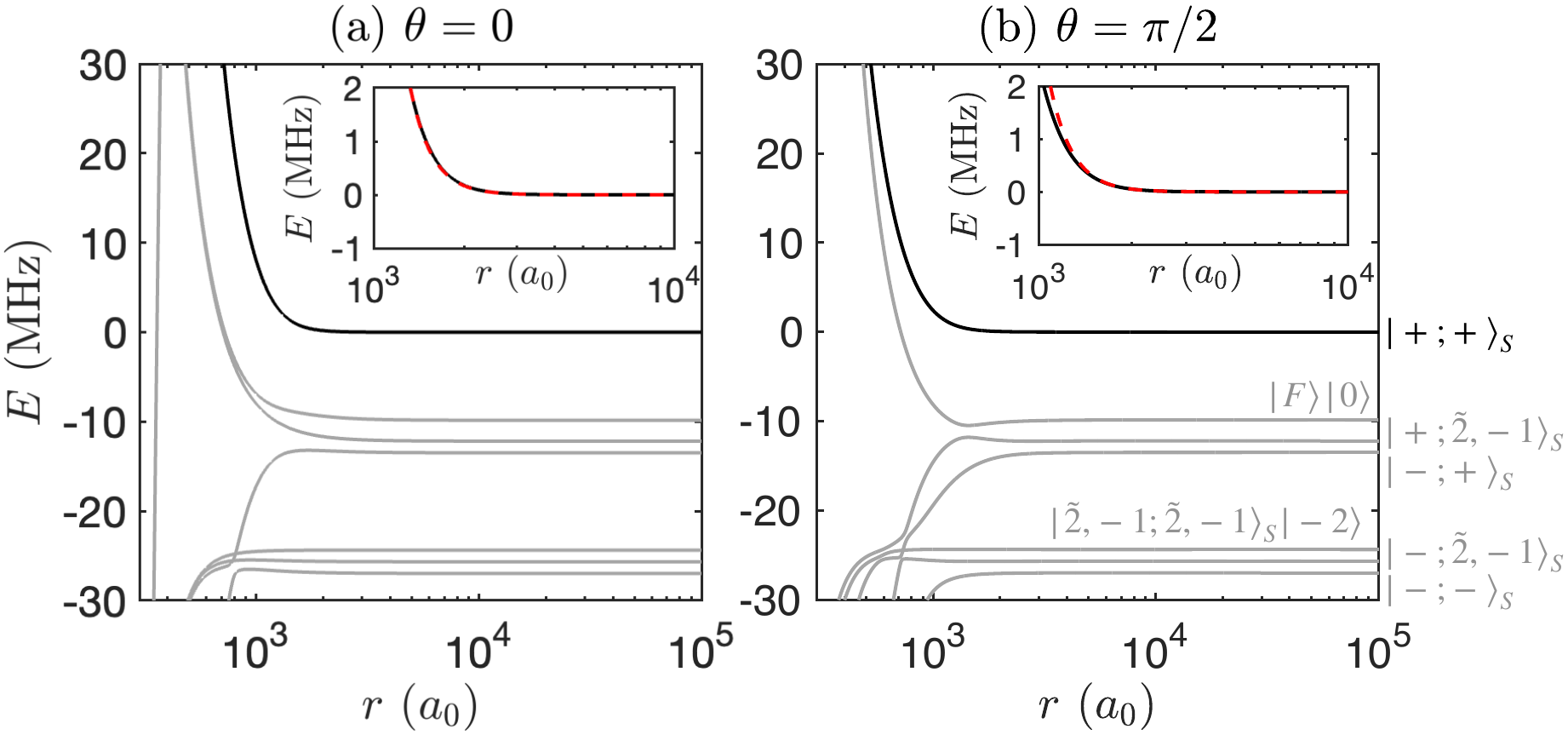}
    \vspace{-20pt}
    \caption{ Adiabatic potential energy curves for two NaCs molecules approaching along (a) $\theta = 0$ and (b) $\theta = \pi/2$ respectively. Parameters are set at compensation with $d{\rm E}_{\rm dc} = 3.25 B_0$, $\Omega/\Delta = 0.733$ and $\Delta = 1.5 \Delta_F$. 
    The shielded adiabatic curve is colored black, while the other inelastic channels are colored gray.  
    Labels for the asymptotic thresholds are ordered from top to bottom by descending energy. 
    The insets are zoomed in versions of their parent plots, comparing the shielded adiabatic curve from full-basis diagonalization (solid black curve) to the effective potential (dashed red curve). }
    \label{fig:compensated_adiabats}
\end{figure}

By inspecting $d_{\rm eff}$ in $V_{\rm eff}(\boldsymbol{r})$, we find that the first-order DDI is fully compensated (i.e. set to zero) at $\Omega = \Omega^* \approx 0.733 \Delta$, showcased in the plot of $d_{\rm eff}^2$ as a function of $\Delta$ and $\Omega$ in subplot (b) of Fig.~\ref{fig:ACDC_shielding_scheme_summary}.  
Notably, increasing or decreasing $\Omega$ away from $\Omega^*$ provides control over the sign of DDI. 
At compensation, there are no attractive first-order DDI along any orientation and, therefore, no FL states. 
By maintaining $\Delta > 0.893 \Delta_F$, we ensure that the $\ket{ \tilde{2}, 0; \tilde{0}, 0 }$ pair rotational state remains energetically closest to the entrance scattering threshold to leverage the F\"orster resonant shielding effect. 
The result is a second-order interaction that is repulsive along all orientations to prevent molecules from entering the short range, showcased in Fig.~\ref{fig:compensated_adiabats} for two interacting ac/dc dressed NaCs molecules at compensation approaching along (a) $\theta = 0$ and (b) $\theta = \pi/2$.  
In these plots, the highest plotted threshold corresponds to the entrance shielded state $\ket{ +; + }_S$ which we offset to zero, followed by $\ket{ F }\ket{ 0 }$ and then other microwave dressed states below those. 
Using the F\"orster resonance in the first excited rotational state allows the shielding barrier to reach a height of $E_{\rm shield} \approx 0.08 B_0$ at around one third the Rabi frequency of the ground state scheme proposed in Ref.~\cite{Gorshkov08_PRL} with $E_{\rm shield} \approx 0.02 B_0$.
As a result, tunneling through the barrier is highly suppressed and two-body loss predominantly results from non-adiabatic transitions into inelastic scattering channels, shown in our calculations below.

\begin{figure}[ht]
    \vspace{-2pt}
    \centering
    \includegraphics[width=\linewidth]{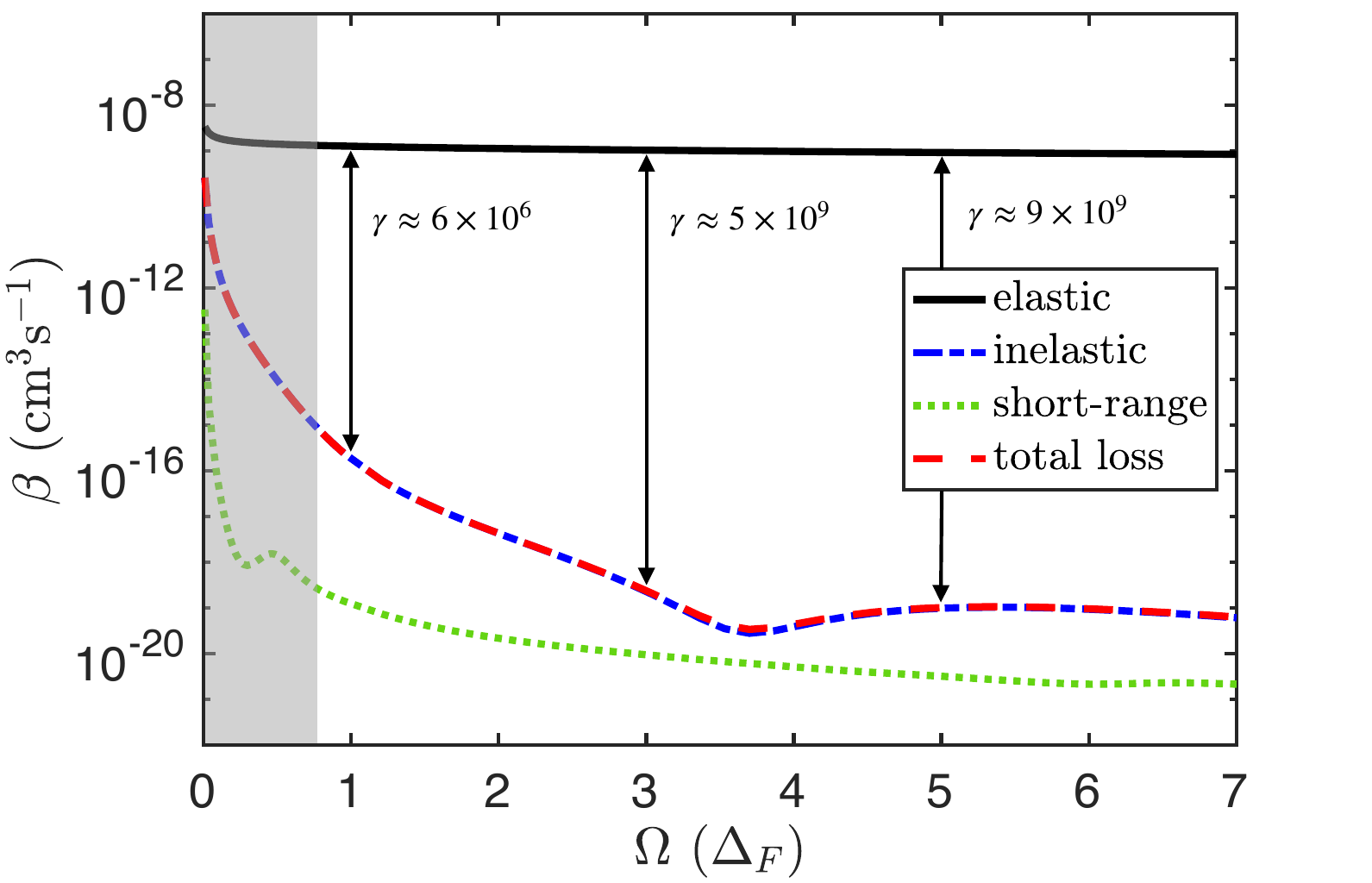}
    \vspace{-20pt}
    \caption{ Calculated two-body elastic (solid black), inelastic (dotted-dashed blue), short-rang (dotted green) and total loss (dashed red) rate coefficients between NaCs molecules at collision energy $E_c/k_B = 100$ nK, as a function of Rabi frequency at compensation with $\Omega = \Omega^*$. The ratio of elastic-to-loss rates are provided at $\Omega/\Delta_F = 1, 3, 5$ in the figure. The gray shaded area indicates the region where shielding is no longer F\"orster resonant dominated with $\Delta < 0.893 \Delta_F$. }
    \label{fig:rateCoefficient_vs_RabiFrequency}
\end{figure}

\textit{Results}--
We determine the efficacy of shielding at compensation through the two-body rate coefficient $\beta$, the product of the integral scattering cross section and the relative collision velocity, obtained from numerical close-coupling scattering calculations \cite{Quemener17_RSC} utilizing an adaptive step-size version of the Johnson log-derivative method \cite{Johnson73_JCP}. 
A universal loss short-range boundary condition \cite{Wang15_NJP} is utilized, that accounts for short-range chemistry \cite{Ni10_Nat} or sticky dynamics \cite{Mayle12_PRA, Bause23_JPCA} both assumed to lead to molecular loss.
The calculations were performed at a $100$ nK collision energy, including rotational states up to $N=4$, photon states $n = 0, -1, -2$, and collisional partial waves up to $L = 8$.
The computed elastic (solid black), inelastic (dashed-dotted blue) and short-range (dotted green) rate coefficients between NaCs molecules are plotted as a function of $\Omega$ in Fig.~\ref{fig:rateCoefficient_vs_RabiFrequency}. 
The high shielding barrier leads to loss that is dominated by inelastic collisions rather than tunneling into the short-range, making the total loss rate coefficient (dashed red curve), comprised of the sum of the former two rate coefficients, almost identical to the inelastic rate coefficient.  
Increasing the Rabi frequency causes the entrance threshold to be further separated from the inelastic scattering channels below, resulting in the observed decreasing trend of the loss rate coefficients to $\sim 10^{-19}$ cm$^3$s$^{-1}$ with increasing $\Omega$. 
Moreover, a slowly varying elastic rate over the range of $\Omega$ plotted in Fig.~\ref{fig:rateCoefficient_vs_RabiFrequency} promises an elastic-to-loss rate coefficient ratio of up to $\gamma \gtrsim 10^{9}$ at large Rabi frequencies, ideal for evaporative cooling into strongly interacting degenerate molecular gases. 
\onecolumngrid

\begin{figure}[ht]
    \centering
    \includegraphics[width=\linewidth]{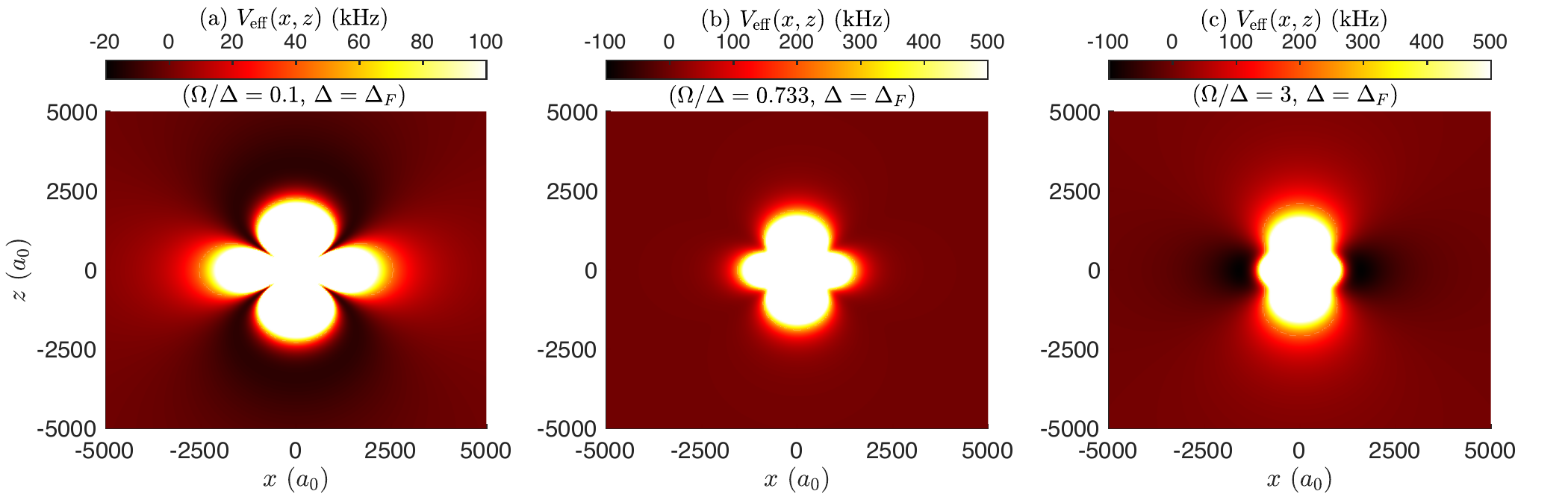}
    \vspace{-20pt}
    \caption{ The effective potential between ac/dc shielded NaCs molecules in the $x,z$-plane, assuming the dipoles are oriented along $\hat{\boldsymbol{z}}$. The subplots are tuned to have (a) dipolar interactions with $\Omega/\Delta = 0.1$; (b) compensated interactions with $\Omega/\Delta = 0.733$; and (c) anti-dipolar interactions with $\Omega/\Delta = 0.3$. The microwaves are detuned by $\Delta = \Delta_F$ in all three plots. The colormaps saturate at 500 kHz except for subplot (a) which saturates at 100 kHz for clarity of the potential wells. }
    \label{fig:tunable_interactions}
\end{figure}

\twocolumngrid

While maintaining F\"orster dominated shielding with $\Delta > 0.893 \Delta_F$, tuning the ratio of $\Omega/\Delta$ provides a control knob between dipolar (head-to-tail attractive) and anti-dipolar (side-to-side attractive) interactions \footnote{ The term anti-dipolar interactions refers to dipole-dipole interactions that have been modified by an overall minus sign. } as showcased in Fig.~\ref{fig:tunable_interactions}. From left to right, the subplots show a change in the character of the effective potential from dipolar, to fully compensated, to anti-dipolar as $\Omega/\Delta$ is tuned across $0.733$ with $\Delta = \Delta_F$. 
Even when tuned away from compensation, shielding still remains effective while allowing a region of $\Omega$ that maintains no FL states in the long-range shielded adiabatic potential.

\begin{figure}[h]
    \centering
    \includegraphics[width=\linewidth]{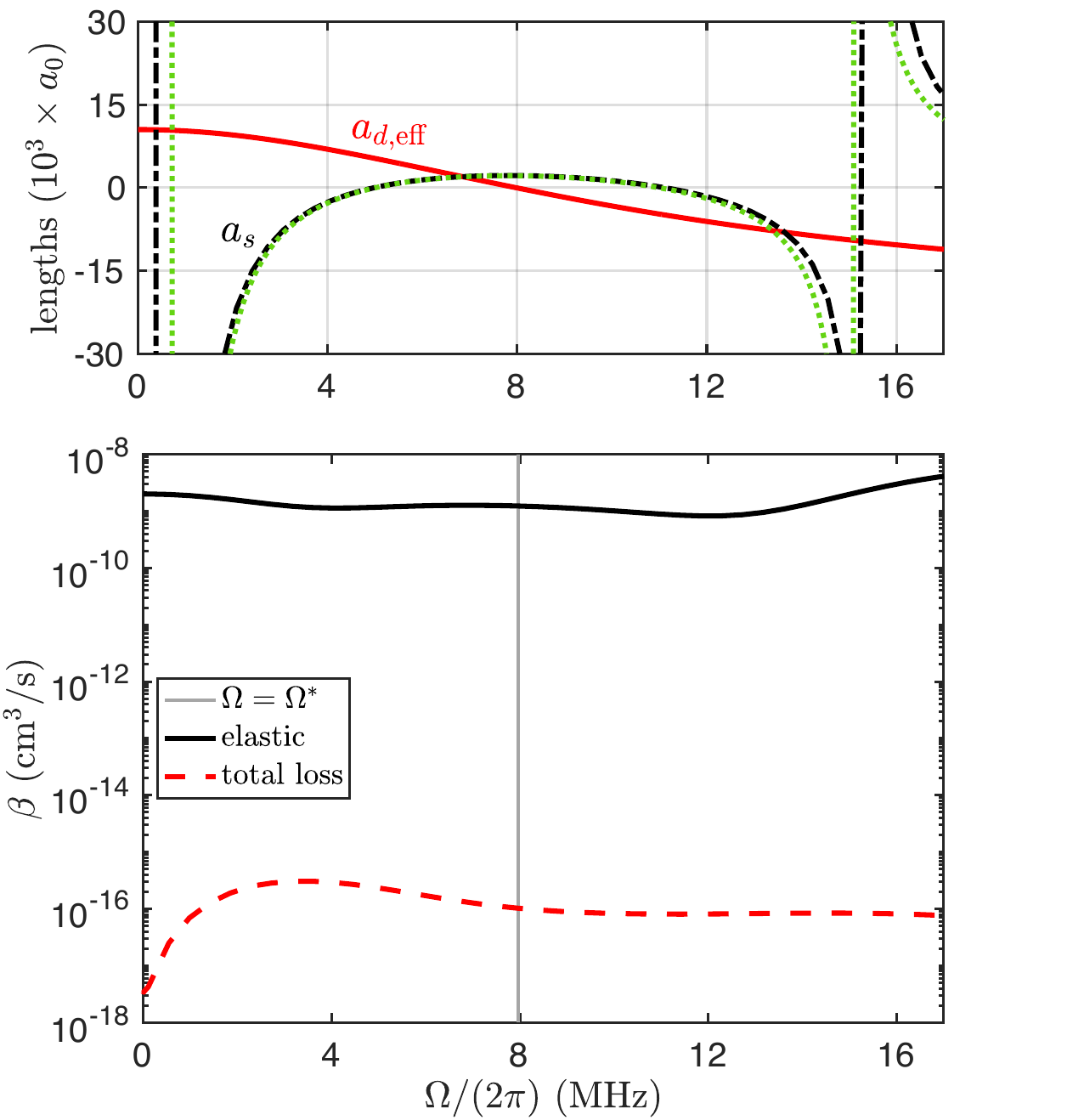}
    \vspace{-20pt}
    \caption{ Upper panel: Effective dipole length (solid red curve), and real part of the $s$-wave scattering length from multichannel calculations (dashed-dotted black curve) and from the effective potential (dotted green curve) as a function of Rabi frequency for NaCs. Lower panel: Calculated two-body elastic (solid black) and total loss (dashed red) rate coefficients at collision energy $E_c/k_B = 100$ nK, as a function of Rabi frequency. 
    The vertical gray line indicates the compensation Rabi frequency $\Omega^*$, at fixed detuning $\Delta = 1.5 \Delta_F$.   }
    \label{fig:rateCoefficient_vs_RabiFrequency_fixedDelta}
\end{figure}

By fixing  $\Delta = 1.5 \Delta_F$ and varying the Rabi frequency, the lower panel of Fig.~\ref{fig:rateCoefficient_vs_RabiFrequency_fixedDelta} shows that $\gamma$ remains $\gtrsim 10^{5}$ within the window without any FL states, with $\Omega^*$ marked by a vertical gray line in the figure. 
The upper panel in Fig.~\ref{fig:rateCoefficient_vs_RabiFrequency_fixedDelta} plots the effective dipole length $a_{d, {\rm eff}} = \mu d_{\rm eff}^2 / (4\pi\epsilon_0\hbar^2)$ (solid red curve) and real part of the $s$-wave scattering length $a_s$ (dashed-dotted black curve) obtained from multichannel close-coupling calculations at $E_c/k_B = 100$ pK, showing two resonant features on either side of $\Omega^*$. We use $\mu$ to denote the reduced mass in the two-molecule center-of-mass frame.  
Each of these features correspond to a threshold resonance as an FL state is pulled from the continuum through either dipolar (left) or anti-dipolar (right) attraction, that are less evident in the collision rate coefficients computed further from threshold at $E_c/k_B = 100$ nK. 
We also plot $a_s$ obtained from scattering calculations on the effective potential of Eq.~(\ref{eq:effective_potential}) as the dotted green curve in the upper panel of Fig.~\ref{fig:rateCoefficient_vs_RabiFrequency_fixedDelta}, showing good agreement with the multichannel calculations.

\textit{Conclusions}--
To summarize, we propose a new protocol for collisionally shielding ultracold molecules with a combination of microwave and static electric fields. 
By utilizing a static field to engineer a F\"orster resonance between two molecules that creates a repulsive intermolecular shield, the microwave couples other pair molecular states to provide control over the strength and character of long-range DDI.
At fields where $d{\rm E}_{\rm dc} = 3.25 B_0$, $\Omega = 0.733 \Delta$ and the detuning exceeds the F\"orster defect by $\Delta > 0.893 \Delta_F$, we find that all FL bound states can be removed from the long-range potential through cancellation of the first-order DDI, while simultaneously maintaining a high shielding efficiency of $\gamma \gtrsim 10^6$ at current day achievable microwave powers.
Changing the ratio of $\Omega/\Delta$ then provides on-demand control of the strength and character of dipolar interactions, alongside the scattering length for utility in quantum simulation \cite{Gadway16_IOP, Bohn17_Science, Cornish24_NatPhys}, studying anisotropic hydrodynamics \cite{Wang22_PRA, Wang22_PRA2, Wang23_PRA, Wang23_PRA2}, and exploring exotic quantum phases \cite{Langen25_PRL, Ciardi25_PRL, Schindewolf25_arxiv}.   
Around the resonant shielding point, $\Delta_F$ has an approximate linear scaling with ${\rm E}_{\rm dc}$ with slope $\partial (\hbar\Delta_F) / \partial (d {\rm E}_{\rm dc}) \approx 0.7$ \cite{Mukherjee25_NJP}. Therefore, ensuring that interactions remain within the bound-state-free region requires that electric field fluctuations only cause variations in $\hbar\Delta_F$ to be $\ll 10^{-3} B_0$. As a conservative estimate, a maximal allowable F\"orster defect fluctuation of $10^{-7} B_0$, would require a dc field stability within $0.7 \times 10^{-7} B_0 / d$, translating to around half a volt per centimeter stability for NaCs molecules.  

For broad applicability, the presented shielding scheme is formulated in an adimensional manner through comparison of external field quantities with intrinsic molecular constants, to ease its implementation on generic diatomic dipolar molecules \cite{GonzalezMartinez17_PRA, Dutta25_PRR}.  
Additionally, although the analysis done in this work focuses on $^{1}\Sigma$ molecules under the rigid-rotor approximation, we expect our shielding scheme to remain robust even for paramagnetic $^2\Sigma$ molecules, since the nuclear and electronic spins can be made spectators to the collision dynamics in F\"orster resonant shielding \cite{Quemener16_PRA, Mukherjee23_PRR}. 
As technology improves to provide stronger microwave sources, our protocol is naturally extended to operate at static electric fields closer to the optimal shielding value ($d{\rm E}_{\rm dc} = 3.4 B_0$), along with correspondingly large microwave detunings and powers. We expect collisional shielding in this regime to be more effective at suppressing loss.    
Rigorous investigations of these claims will be explored in a subsequent work.

\begin{acknowledgments}

The author acknowledges inspiring discussions with T. Karman and J. L. Bohn, and helpful technical remarks from C. Nunez and J. Luke. 
The author thanks H. R. Sadeghpour for a careful reading of the manuscript.   
This project was supported by ITAMP with funding from the National Science Foundation.  

\end{acknowledgments}

\appendix

\section{ Derivation of the ac/dc effective shielding potential \label{app:effective_potential_derivation} }

The Hamiltonian for a single rigid rotor molecule dressed by d.c and ac fields is given as
\begin{align}
    {\cal H}
    &=
    {\cal H}_{\rm rot} 
    +
    {\cal H}_{\rm dc}
    +
    {\cal H}_{\rm mw}
    +
    {\cal H}_{\rm ac} \\
    &=
    B_{\rm rot} \boldsymbol{N}^2
    -
    \boldsymbol{d} \cdot \boldsymbol{{\rm E}}_{\rm dc}
    +
    \hbar \omega_{\sigma^{+}}
    \left(
    a_{\sigma^{+}}^{\dagger} a_{\sigma^{+}} - n_{0}
    \right) \nonumber\\
    &\quad 
    -
    \frac{ {\rm E}_{\rm ac} }{ 2 \sqrt{ n_{0} } }
    \left(
    d_{+}^{\dagger} a_{\sigma^{+}}^{\dagger} 
    +
    d_{+} a_{\sigma^{+}}
    \right), \nonumber
\end{align}
where $n_{0}$ is the background number of microwave photons, $a_{\sigma^{+}}^{\dagger}$ ($a_{\sigma^{+}}$) is the creation (annihilation) operator of $\sigma^+$ polarized microwave photons, and $d_{+} = -(d_x + i d_y)/\sqrt{2}$ is a rank-1 spherical dipole tensor.
The matrix elements of the dc Stark Hamiltonian are:
\begin{align}
    \bra{ N', m' }
    \mathcal{H}_{\rm dc}
    & \ket{ N, m }
    =
    -d {\rm E}_{\rm dc} ( -1 )^{m'} \nonumber\\
    &\times
    \sqrt{ (2 N + 1) (2 N' + 1) }  \nonumber\\
    &\times
    \begin{pmatrix}
        N' & 1 & N \\
        0 & 0 & 0
    \end{pmatrix}
    \begin{pmatrix}
        N' & 1 & N \\
        -m' & 0 & m
    \end{pmatrix}.
\end{align} 
The static field-dressed rotational states $|{ \tilde{N}, m }\rangle$, are then linear combinations of the bare rotational states that are eigenstates of $({\cal H}_{\rm rot} + {\cal H}_{\rm dc}) |{ \tilde{N}, m }\rangle = {\varepsilon}_{|\tilde{N}, m\rangle} |{ \tilde{N}, m }\rangle$. 

A microwave then drives the $|{ \tilde{N}, m }\rangle = |{ \tilde{1}, 0 }\rangle \rightarrow |{ \tilde{2}, +1 }\rangle$ dc dressed state transition, which in the joint molecule-photon basis $|{ \tilde{N}, m }\rangle\ket{ n }$ of just the two driven states, the Hamiltonian takes the matrix representation:
\begin{align}
    \boldsymbol{{\cal H}}
    &=
    \begin{pmatrix}
        0 & \hbar\Omega/2 \\
        \hbar\Omega/2 & -\hbar\Delta
    \end{pmatrix},
\end{align}
where the Rabi frequency is given by
\begin{align}
    \hbar\Omega 
    &= 
    d {\rm E}_{\rm ac}
    \sum_{ N', N = 0 }^{\infty}
    \langle{ \tilde{2}, +1 }| N', +1 \rangle
    \langle{ N, 0 }| \tilde{1}, 0 \rangle  \nonumber\\
    &\quad\quad\quad\quad
    \quad\quad \times 
    \sqrt{ \frac{ (N+1)(N+2) }{ 8 N (N+2) + 6 } },
\end{align}
along with detuning $\hbar\Delta = \hbar\omega_{\sigma^{+}} - ( {\varepsilon}_{|\tilde{2}, +1\rangle} - {\varepsilon}_{|\tilde{1}, 0\rangle} )$. 
The $2 \times 2$ Hamiltonian of states bright to the microwaves above is easily diagonalized to give the ac dressed states
\begin{subequations} 
\begin{align}
    \ket{ - }
    &=
    \cos\varphi |{ \tilde{1}, 0}\rangle|{ 0 }\rangle
    -
    \sin\varphi |{ \tilde{2}, +1}\rangle|{ -1 }\rangle, \\
    \ket{ + }
    &=
    \sin\varphi |{ \tilde{1}, 0}\rangle|{ 0 }\rangle
    +
    \cos\varphi |{ \tilde{2}, +1}\rangle|{ -1 }\rangle, 
\end{align}
\end{subequations}
with corresponding energies 
\begin{subequations}
\begin{align}
    \varepsilon_{-}
    &=
    -\frac{ \hbar\Delta }{ 2 }
    -
    \frac{ \hbar }{ 2 }
    \sqrt{ \Delta^2 + \Omega^2 }, \\
    \varepsilon_{+}
    &=
    -
    \frac{ \hbar\Delta  }{ 2 }
    +
    \frac{ \hbar }{ 2 }
    \sqrt{ \Delta^2 + \Omega^2 }, 
\end{align}
\end{subequations}
where we defined the mixing angle
\begin{align}
    \varphi
    &=
    \tan^{-1}
    \left(
    \frac{ \Delta }{ \Omega }
    +
    \sqrt{ 1 + \left( \frac{ \Delta }{ \Omega } \right)^2 }
    \right). 
\end{align}

Considering two-molecules, we utilize the truncated basis of Eq.~(\ref{eq:DCdressed_symmetrized_basis}) to get the non-interacting Hamiltonian
\begin{align}
    & \boldsymbol{{\cal H}} = \\
    &
    \begin{pmatrix}
        0 & 0 & \hbar{\Omega}/\sqrt{2} & 0 & 0 & 0 \\ 
        0 & -\hbar\Delta_F & 0 & 0 & 0 & 0 \\
        \hbar{\Omega}/\sqrt{2} & 0 & -\hbar\Delta & 0 & \hbar{\Omega}/2 & 0 \\
        0 & 0 & 0 & -\hbar\Delta & 0 & \hbar{\Omega}/\sqrt{2} \\
        0 & 0 & \hbar{\Omega}/2 & 0 & -2 \hbar\Delta & 0 \\
        0 & 0 & 0 & \hbar{\Omega}/\sqrt{2} & 0 & -2 \hbar\Delta
    \end{pmatrix}. \nonumber
\end{align}
Diagonalizing the matrix above gives the ac/dc dressed eigenstates and eigenenergies:
\begin{align} 
    \begin{tabular}{l|c}
        dressed state & energy \\
        \hline
        $|{ +; + }\rangle_S$ & $2 \varepsilon_{+}$ \\
        $|{ F }\rangle\ket{ 0 }$ & $-\hbar\Delta_{F}$ \\
        $|{ +; \tilde{2}, -1 }\rangle_S$ & $\varepsilon_{+} - \hbar\Delta$ \\
        $|{ -; + }\rangle_S$ & $\varepsilon_{-} + \varepsilon_{+}$ \\
        $|{ -; \tilde{2}, -1 }\rangle_S$ & $\varepsilon_{-} - \hbar\Delta$ \\
        $|{ -; - }\rangle_S$ & $2 \varepsilon_{-}$
    \end{tabular}
\end{align}
with the state representation for the two three states above is represented in the basis of Eq.~(\ref{eq:DCdressed_symmetrized_basis}) by
\begin{subequations}
\begin{align}
    |{ +; + }\rangle_S
    &=
    \frac{ 1 }{ 2 \sqrt{\left(\delta ^2 + 1\right) \left[ 1 + 2 \delta  \left(\sqrt{\delta ^2+1}+\delta \right) \right] } } \nonumber\\
    &\quad \times
    \begin{pmatrix}
        2 \delta  \left(\sqrt{\delta ^2+1}+\delta \right)+1 \\
        0 \\
        \sqrt{2} \left(\sqrt{\delta ^2+1}+\delta \right) \\
        0 \\
        1 \\
        0
    \end{pmatrix},
\end{align}
\end{subequations}
where $\delta = \Delta/\Omega$.

We take that at long-range, the molecules primarily interact through dipole-dipole interactions:
\begin{align}
    V_{\rm dd}
    &=
    -\frac{ \sqrt{ 6 } }{ 4 \pi \epsilon_0 r^3 }
    \sum_{p=-2}^2
    (-1)^p C_{2, -p}(\theta, \phi)
    [ \boldsymbol{d}_A \otimes \boldsymbol{d}_B ]_p^{(2)},
\end{align}
where $C_{\ell, m}(\theta, \phi)$ are reduced spherical harmonics. 
The dipole-dipole interaction matrix elements in the bare rotational basis are computed as
\begin{align} \label{eq:DDI_matrix_elements}
    & \langle N'_A, m'_{A}; N'_B, m'_{B} |
    V_{\rm dd}(\boldsymbol{r})
    | N_A, m_{A}; N_B, m_{B} \rangle
    \nonumber\\
    &=
    -\frac{ 2 C_{2,-q}(\theta, \phi) }{ 4 \pi \epsilon_0 r^3 } 
    (-1)^{m'_{A} + m'_{B}} 
    \frac{ \sqrt{ 30 } }{ 2 }
    \begin{pmatrix}
        1 & 1 & 2 \\
        q_A & q_B & -q
    \end{pmatrix} \nonumber\\
    &\quad \times
    d \sqrt{ (2 N'_A + 1) (2 {N_A} + 1) } \nonumber\\
    &\quad\quad \times
    \begin{pmatrix}
        N'_A & 1 & {N_A} \\
        -m'_{A} & q_A & m_{A}
    \end{pmatrix}
    \begin{pmatrix}
        N'_A & 1 & {N_A} \\
        0 & 0 & 0
    \end{pmatrix} \nonumber\\
    &\quad \times
    d \sqrt{ (2 N'_B + 1) (2 {N_B} + 1) } \nonumber\\
    &\quad\quad \times
    \begin{pmatrix}
        N'_B & 1 & {N_B} \\
        -m'_{B} & q_B & m_{B}
    \end{pmatrix}
    \begin{pmatrix}
        N'_B & 1 & {N_B} \\
        0 & 0 & 0
    \end{pmatrix},
\end{align}
where $q_A = m'_{A} - m_{A}$, $q_B = m'_{B} - m_{B}$ and $q = q_A + q_B$ are angular momentum transfer quantum numbers, and $C_{\ell, m_{\ell}}(\theta, \phi)$ are reduced spherical harmonics. 
We would, however, like to interpret these matrix elements in the dc dressed basis, which can be done by performing a change of basis from the bare to static field-dressed states, while noting the conservation of magnetic quantum numbers.
Moreover, the relevant matrix elements are those in the pair molecular symmetrized sector, with symmetrization
\begin{align}
    | {A; B} \rangle_S
    &=
    \frac{ | {A; B} \rangle_S + | {B; A} \rangle_S }{ \sqrt{ 2 ( 1 + \delta_{\tilde{N}_A, \tilde{N}_B} \delta_{m_A, m_B} ) } }
\end{align}
having employed the short-hand notation $|{A; B}\rangle = |{ \tilde{N}_A, m_{A}; \tilde{N}_B, m_{B} }\rangle$.  
For convenience, it is useful to introduce the effective direct and induced dipole-moments
defining
\begin{subequations}
\begin{align}
    d_g 
    &=
    d_{\tilde{1}}^{(0)}, \quad
    d_{e}
    =
    d_{\tilde{2}}^{(+1)}, \quad
    d_{\bar{e}}
    =
    d_{\tilde{2}}^{(-1)}, \\
    d_{g\rightarrow\bar{e}}
    &=
    d_{\tilde{1}\rightarrow\tilde{2}}^{0\rightarrow -1}, \quad
    d_{\bar{e}\rightarrow{g}}
    =
    d_{\tilde{2}\rightarrow\tilde{1}}^{-1\rightarrow 0}, \\
    d_{g\rightarrow{e}}
    &=
    d_{\tilde{1}\rightarrow\tilde{2}}^{0\rightarrow +1}, \quad
    d_{e\rightarrow{g}}
    =
    d_{\tilde{2}\rightarrow\tilde{1}}^{+1\rightarrow 0}, \\
    d_{l}
    &=
    d_{\tilde{0}}^{(0)}, \quad
    d_{u}
    =
    d_{\tilde{2}}^{(0)}, \\
    d_{l\rightarrow{u}}
    &=
    d_{\tilde{0}\rightarrow\tilde{2}}^{0\rightarrow 0}, \quad 
    d_{u\rightarrow{l}}
    =
    d_{\tilde{2}\rightarrow\tilde{0}}^{0\rightarrow 0},
\end{align}
\end{subequations}
where the explicit dipole matrix elements are
\begin{subequations}
\begin{align}
    d_{\tilde{N}}^{(m)}
    &=
    d 
    \sum_{ N', N }
    \langle \tilde{N}, m | N', m \rangle
    \langle N, m | \tilde{N}_, m \rangle \nonumber\\
    &\quad\quad\quad\quad \times
    \sqrt{ (2 N' + 1) (2 {N} + 1) } \nonumber\\
    &\quad\quad\quad\quad \times
    \begin{pmatrix}
        N' & 1 & {N} \\
        -m & 0 & m
    \end{pmatrix}
    \begin{pmatrix}
        N' & 1 & {N} \\
        0 & 0 & 0
    \end{pmatrix}, \\
    d_{\tilde{N}\rightarrow\tilde{N}'}^{ m \rightarrow m' }
    &=
    d 
    \sum_{ N', N }
    \langle \tilde{N}', m' | N', m' \rangle
    \langle N, m | \tilde{N}_, m \rangle \nonumber\\
    &\quad\quad\quad\quad \times
    \sqrt{ (2 N'_B + 1) (2 {N_B} + 1) } \nonumber\\
    &\quad\quad\quad\quad \times
    \begin{pmatrix}
        N' & 1 & {N} \\
        -m' & \Delta{m} & m
    \end{pmatrix}
    \begin{pmatrix}
        N' & 1 & {N} \\
        0 & 0 & 0
    \end{pmatrix}.
\end{align}
\end{subequations}
with $\Delta{m} = \pm \sigma$, defined without the $(-1)^{m'}$ phase factor.

Now computing the dipole-dipole interaction matrix elements in the symmetrized dc-dressed basis of Eq.~(\ref{eq:DCdressed_symmetrized_basis}), the diagonal matrix elements are 
\begin{subequations}
\begin{align}
    V_{gg}
    &=
    \frac{ d_g^2 }{ 4 \pi \epsilon_0 }
    \frac{ 1 - 3 \cos^2\theta }{ r^3 },
    \\
    V_{\bar{e}g}
    &=
    -\frac{ d_g d_{\bar{e}} + d_{g\rightarrow\bar{e}} d_{\bar{e}\rightarrow{g}} / 2 }{ 4 \pi \epsilon_0 }
    \frac{ 1 - 3 \cos^2\theta }{ r^3 },
    \\
    V_{eg}
    &=
    -\frac{ d_{g} d_{e} + d_{g\rightarrow{e}} d_{{e}\rightarrow{g}} / 2 }{ 4 \pi \epsilon_0 }
    \frac{ 1 - 3 \cos^2\theta }{ r^3 },
    \\
    V_{e\bar{e}}
    &=
    \frac{ d_{\bar{e}} d_{e} }{ 4 \pi \epsilon_0 }
    \frac{ 1 - 3 \cos^2\theta }{ r^3 },
    \\
    V_{ee}
    &=
    \frac{ d_{e}^2 }{ 4 \pi \epsilon_0 }
    \frac{ 1 - 3 \cos^2\theta }{ r^3 },
    \\
    V_{F}
    &=
    \frac{ d_{l} d_{u} + d_{l\rightarrow{u}} d_{u\rightarrow{l}} }{ 4 \pi \epsilon_0 }
    \frac{ 1 - 3 \cos^2\theta }{ r^3 },
\end{align}
\end{subequations}
utilizing the notation $V_{ab} = \langle a; b | V_{\rm dd} | a; b \rangle$ with $a,b \in \{ g, e, \bar{e}, F \}$.
Similarly, the non-trivial off-diagonal elements are
\begin{subequations}
\begin{align}
    U = 
    \langle \bar{e}; g |
    V_{\rm dd}
    | e; g \rangle
    &=
    -\frac{ \sqrt{6} d_{g \rightarrow e} d_{\bar{e} \rightarrow g} }{ 4 \pi \epsilon_0 }
    \frac{ C_{2,+2} }{ r^3 }, \\
    W = 
    \langle g; g |
    V_{\rm dd}
    | F \rangle
    &=
    \langle F |
    V_{\rm dd}
    | g; g \rangle \\
    &=
    -\frac{ \sqrt{ 2 } d_{\tilde{1}\rightarrow\tilde{2}}^{0\rightarrow 0} d_{\tilde{1}\rightarrow\tilde{0}}^{0\rightarrow 0} }{ 4 \pi \epsilon_0 }
    \frac{ 1 - 3 \cos^2\theta }{ r^3 }, \nonumber
\end{align}
\end{subequations}
which allow us to write the dipole-dipole coupling matrix elements as
\begin{align}
    \boldsymbol{V}_{\rm dd}
    &=
    \begin{pmatrix} 
        V_{gg} & W & 0 & 0 & 0 & 0 \\
        W & V_{F} & 0 & 0 & 0 & 0 \\
        0 & 0 & V_{\bar{e}g} & U & 0 & 0 \\
        0 & 0 & U^* & V_{eg} & 0 & 0 \\
        0 & 0 & 0 & 0 & V_{e\bar{e}} & 0 \\
        0 & 0 & 0 & 0 & 0 & V_{ee}
    \end{pmatrix}.
\end{align}
The first-order interactions with state $\ket{ +;+ }_S$ is computed as
\begin{subequations}
\begin{align}
    V_{\rm dd}^{(1)}(\boldsymbol{r})
    &=
    _S\bra{ +;+ }
    V_{\rm dd}(\boldsymbol{r}) 
    \ket{ +;+ }_S \\
    &=
    \frac{ d_{\rm eff}^2 }{ 4 \pi \epsilon_0 }
    \frac{ 1 - 3 \cos^2\theta }{ r^3 }, \nonumber\\
    d_{\rm eff}^2 
    &=
    ( \alpha_{-} d_{e} - \alpha_{+} d_g )^2 - \alpha_{-} \alpha_{+} d_{g\rightarrow{e}} d_{{e}\rightarrow{g}},
\end{align}
\end{subequations}
where $\alpha_{\pm} = \{ [ 1 + 2 \delta \left(\delta \pm \sqrt{\delta^2 + 1}\right) ] / ( 4\delta^2 + 4 ) \}^{1/2}$, $\delta = \Delta/\Omega$ and $d_{\rm eff}^2$ is the squared effective dipole moment. 
To obtain the second-order interactions, we consider only the two energetically closest states and utilize second-order perturbation theory to get 
\begin{align}
    V^{(2)}_{\rm dd}(\boldsymbol{r})
    &\approx 
    \frac{ | \bra{0}\bra{ F } {V}_{\rm dd}(\boldsymbol{r}) \ket{ +; + }_S |^2 }{ 2\varepsilon_{+} - ( - \hbar\Delta_F ) } \nonumber\\
    &\quad
    +
    \frac{ | _S\bra{ +; \tilde{2}, -1 } {V}_{\rm dd}(\boldsymbol{r}) \ket{ +; + }_S |^2 }{ 2\varepsilon_{+} - ( \varepsilon_{+} - \hbar\Delta ) } \nonumber\\
    &=
    \frac{ \alpha_{+}^2 | W(\boldsymbol{r}) |^2 }{ 2\varepsilon_{+} + \hbar\Delta_F }
    +
    \frac{ \alpha_+^2 - \alpha_+ \alpha_- }{ 2 \delta \sqrt{ 1 + \delta^2 } }
    \frac{ |U(\boldsymbol{r})|^2 }{ \varepsilon_+ + \hbar\Delta }.
\end{align}
Combining the first- and second-order potentials above then gives the effective potential in Eq.~(\ref{eq:effective_potential}) and (\ref{eq:secondorder_potential}).

\bibliography{main.bib} 

\end{document}